\documentclass[a4paper,twoside]{article}

\usepackage{natbib}
\usepackage{graphicx}
\usepackage{verbatim}
\usepackage{hyperref}
\usepackage{caption}
\usepackage{subcaption}
\usepackage{epsfig}
\usepackage{calc}
\usepackage{amssymb}
\usepackage{amstext}
\usepackage{amsmath}
\usepackage{amsthm}
\usepackage{multicol}
\usepackage{pslatex}
\usepackage{apalike}
\usepackage{algorithm2e}
\usepackage[bottom]{footmisc}
\usepackage{SCITEPRESS}     

\begin{document}
\title{Decentralized Privacy-Preserving Federal Learning of Computer Vision Models on Edge Devices}

\author{
\authorname{Damian Harenčák\sup{1}, Lukáš Gajdošech\sup{1}\sup{2}\orcidAuthor{0000-0002-8646-2147}, Martin Madaras\sup{1}\sup{2}\orcidAuthor{0000-0003-3917-4510}}
\affiliation{\sup{1}Faculty of Mathematics, Physics and Informatics, Comenius University, Bratislava, Slovakia}
\affiliation{\sup{2}Skeletex Research, Slovakia}
\email{gajdosech@fmph.uniba.sk, madaras@skeletex.xyz}
}


\abstract{
Collaborative training of a machine learning model comes with a risk of sharing sensitive or private data. Federated learning offers a way of collectively training a single global model without the need to share client data, by sharing only the updated parameters from each client's local model. A central server is then used to aggregate parameters from all clients and redistribute the aggregated model back to the clients. Recent findings have shown that even in this scenario, private data can be reconstructed only using information about model parameters. Current efforts to mitigate this are mainly focused on reducing privacy risks on the server side, assuming that other clients will not act maliciously. In this work, we analyzed various methods for improving the privacy of client data concerning both the server and other clients for neural networks. Some of these methods include homomorphic encryption, gradient compression, gradient noising, and discussion on possible usage of modified federated learning systems such as split learning, swarm learning or fully encrypted models. We have analyzed the negative effects of gradient compression and gradient noising on the accuracy of convolutional neural networks used for classification. We have shown the difficulty of data reconstruction in the case of segmentation networks. We have also implemented a proof of concept on the NVIDIA Jetson TX2 module used in edge devices and simulated a federated learning process. 
}

\onecolumn \maketitle \normalsize \setcounter{footnote}{0} \vfill

\section{Introduction}
Rapid advancements in machine learning have enabled powerful predictive models across a wide range of applications, but they often depend on access to large centralized datasets. In many industrial scenarios, however, aggregating sensitive data into a single repository is neither feasible nor permissible due to privacy, security, and regulatory constraints. A common alternative is federated learning (FL), a 
paradigm in which clients retain raw data locally and only exchange model updates, such as gradients or parameter deltas, with a coordinating server \cite{yang2020federated, openFL, GAIN, HEPPFL}. While FL mitigates some privacy risks by avoiding direct sharing of raw data, it does not guarantee complete confidentiality. Under certain conditions, adversaries can reconstruct client data from shared updates (DLG algorithm \cite{GLeakage}), raising new security concerns for FL systems. The focus is given on environments typical for 3D computer vision edge devices, including the assumption of hardware with limited resources and the need for lightweight solutions capable of running on this hardware. All our implementations and experiments are available on GitHub\footnote{\href{http://skeletex.xyz/redirect/visapp2026fl}{http://skeletex.xyz/redirect/visapp2026fl}
}.

\subsection{Motivation}
Clients demand increasingly accurate statistical models to process their data, e.g. quality inspection in industry. 
Proprietary scan data often contains competitive secrets that cannot be shared outside of the local manufacturing network. Privacy-preserving FL offers a solution: by keeping raw data on-site and only sharing encrypted or obfuscated model updates, clients can locally train and share robust models without exposing sensitive scan data. This motivates the development of techniques such as homomorphic encryption, gradient compression, noise injection, and alternative model topologies to ensure data confidentiality throughout the decentralized training process. 

\section{Theoretical Background}

This part provides a basic definition of techniques used in decentralized training. 


\subsection{Federated Learning}
Federated learning (FL) \cite{mcmahanFederatedLearning, yang2020federated} is a collaborative training approach in which multiple participants, also called clients, jointly train a single shared global model, such as a convolutional neural network, without sharing their raw training data. Instead, clients transmit model updates (e.g., parameters or gradients) to a central server, which aggregates these updates to improve the global model. This method is designed to operate in environments where the data across clients is not independent and identically distributed, 
meaning that the local data on each client may not fully represent the overall population distribution. The primary motivation behind FL is to address challenges related to data privacy, data minimization, and data access rights. 

\subsubsection{Centralized federated learning}
In centralized FL, a central server is used to aggregate model updates. This server is also responsible for selecting which nodes participate in training and for coordinating the overall learning process. It manages the scheduling of training rounds and the communication between nodes, ensuring that the updates are systematically merged into a shared global model. Centralizing all these operations within one server introduces the risk of a single point of failure in the system and requires a certain level of trust from all clients.

\subsubsection{Decentralized federated learning}
In decentralized FL, no single entity dictates the training process. Instead, individual nodes work together to decide which peers join each training round, and to coordinate the scheduling of training sessions to obtain a global model. Decentralization eliminates the risk of single point of failure and can benefit from a transparent history of updates, trusted across all clients.


\subsubsection{Federated averaging algorithm}
McMahan et al. \cite{mcmahanFederatedLearning} have proposed an algorithm called Federated averaging (FedAvg) aimed for a centralized version of FL, consisting of several steps and describing the entire FL training process:
\begin{enumerate}
    \item Initialization - the process starts by initializing the global model parameters. 
    \item Selection of clients - next, a subset of clients is chosen from the available pool. 
    \item Local training - each selected client receives a copy of the global model parameters and proceeds to update its local model. The model is then trained using the client's local data.
    \item Aggregation - upon completing local training, each client sends its updated model parameters back to the central server. The server aggregates these updates by computing a weighted average of the parameters using the formula:
    $$
    w_{i+1} = \sum_{k=1}^N \frac{n_k}{m_i} w_{i+1}^k,
    $$
    \noindent where $N$ is the number of clients, $n_k$ represents the number of data points used by $k$-th client, $m_i = \sum_{k=1}^Nn_k$ is the total number of data points, and $w_{i+1}^{k}$ are the local parameters updated by the $k$-th client.
    \item Repeat - steps 
    are repeated until the global model achieves the desired level of convergence.
\end{enumerate}

\subsubsection{Privacy challenges}
In FL, raw data never leaves the client's side. This, however, does not guarantee absolute privacy. Even when only gradients or updated model parameters are shared, there is still a risk that the original data can be reconstructed. Ligeng Zhu et al. \cite{GLeakage} highlight the vulnerabilities in traditional gradient-sharing mechanisms by demonstrating "Deep Leakage from Gradients" (DLG). The study empirically validates that training data can be reverse-engineered from shared gradients, raising significant privacy concerns.

Authors of DLG highlight some protective mechanisms that can be used for this mitigation. Gradient compression \cite{GCompression}, where we prune small values, provided good results in their experiments,
when considering the accuracy-security trade-off. Other examined methods include differential privacy (adding noise to the gradients), larger batch sizes, or various encryption schemes.

\subsubsection{Bits of Security}

The "bits of security" metric is a widely used standard for quantifying the computational effort required to break a cryptographic encryption. Specifically, a cipher is said to provide "$n$ bits of security" if the best known attack against it would require approximately $2^n$ operations to succeed. The bits of security estimate assume that the adversary is limited to known algorithms.
Modern cryptographic standards, such as NIST, ENISA or ISO, recommend that systems deployed today should aim to provide at least 112-128 bits of security.

\subsection{Methods for Server-side privacy}

\subsubsection{Homomorphic encryption}

In FL, a standard method for mitigating the risk of gradient leakage from a central server is the application of specialized encryption schemes \cite{GAIN}, called homomorphic encryption, which facilitates secure aggregation of gradients. Additively homomorphic encryption (AHE) allows arithmetic addition to be performed directly on encrypted data. Formally, if  $E(p)$  denotes the encryption function, $D(c)$  the decryption function and $ADD(c_1,c_2)$ the encrypted addition function, the property can be stated as follows:

$$D\big(ADD(E(p_1),E(p_2))\big) = p_1 + p_2,$$

\noindent where $p_1,p_2\in \mathbb{R}$ are plaintext inputs. This property ensures that the sum of encrypted values after decryption equals the sum of the original plaintexts.
By enabling the addition of ciphertexts without requiring decryption, homomorphic encryption allows gradients to be encrypted before transmission. The server can then aggregate these encrypted gradients, and only the aggregated result is decrypted once sent back to clients.

Broader class, generally called fully homomorphic encryption (FHE
), also has the property of encrypted multiplication. Formally, 
if $MUL(c_1,c_2)$ denotes the encrypted multiplication function, the property can be stated as follows:

$$D\big(MUL(E(p_1),E(p_2))\big) = p_1 * p_2.$$

\noindent 
The Paillier and CKKS encryption schemes are among the widely used 
in the context of  FL\cite{HEPPFL, GAIN, chunkCNN}.

\subsubsection{Paillier encryption}
Paillier encryption \cite{Paillier} is an AHE encryption scheme designed for integers that also supports scalar multiplication.
Let $a$ and $b$ denote two $n$-bit primes, let $N=ab$, then the public key $PK=N$ and secret key $SK=(N,\phi(N))$ , where $\phi(N)=(a-1)(b-1)$. If we have message $m\in Z_N$, we can obtain the encrypted ciphertext:
$$c=(1+N)^m\cdot r^N \; mod \; N^2,$$

\noindent where $r\in Z_N$ is a random integer s.t. $0<r<N$ and $gcd(r,N)=1$. The decryption algorithm decrypts by computing: \vspace{-3mm}
$$m=\frac{(c^{\phi(N)}\; mod \; N^2)-1}{N}\cdot \phi(N)^{-1} \; mod \; N,$$

\noindent where the fraction represents floor division\footnote{$\frac{a}{b}=v$ where $v\geq0$ is the largest integer s.t. $a\geq vb$}.


Paillier encrypts each scalar individually. In the context of FL, this means that if we want to secure an entire gradient vector, we must encrypt each element one by one. 
It is designed for exact arithmetic operations over integers. 

The security of this encryption primarily depends on the bit size of the key\footnote{This is the bit size of public key $N$, computed as the product of two large  $n$-bit primes}, where a larger key size enhances security but increases computational cost and the size of ciphertexts. Recommended size for practical purposes is 2048 bits \cite{ECDSAPaillierSecurityLevels}, corresponding to 112 bits of security, at minimum. Smaller sizes are nowadays considered not secure enough.

\subsubsection{CKKS encryption}
CKKS encryption \cite{CKKS} is a FHE encryption scheme designed for floating-point numbers that also supports scalar multiplication of plaintext (vector). The details of the inner workings of this scheme are not the subject of this work, so we only look at defining characteristics, mainly in comparison to Paillier encryption.

CKKS scheme supports the encryption of an entire vector in a single ciphertext through a process known as packing. This allows us to perform operations on many numbers simultaneously, which can lead to significant overhead improvements.
It is based on approximate arithmetic on real or complex numbers and operates within polynomial rings of algebraic integers, which provide a space for encoding and computing on vectors. 
Each operation on an encrypted vector adds noise to the result, which can possibly accumulate to a significant precision error, primarily when using multiplications. This scheme involves several adjustable parameters:
\begin{itemize}
\item polynomial degree: degree of polynomial that defines the ring. Higher degrees generally increase security but also the computational load,
\item ciphertext modulus: a larger modulus can support more complex computations before noise accumulates significantly but may reduce performance,
\item scaling factor: a constant used to convert plaintext numbers into integer representation before encryption, and then to recover approximate results after computations. It helps manage precision and control noise during operations.
\end{itemize}

\subsection{Method for Client-side Privacy}

\subsubsection{Gradient compression}
Gradient compression \cite{GCompression} removes (sets to zero) "insignificant" values from the gradient. Let $G\in \mathbb{R^n}=(g_1, g_2, \dots, g_n)$ be a vector that represents the flattened gradient obtained from training a neural network model. Let $C(G, \epsilon)$ be a compression function where  $\epsilon > 0$ is a small constant. We define $C$ as following:
$$
C(G,\epsilon) = (g'_1, \dots, g'_n) \quad{ where }\quad g'_i=\begin{cases}
g_i, & \text{if } |g_i| > \epsilon \\
0, & \text{else}
\end{cases}.
$$
Since $G$ is dependent on specific network structure and training samples, we will instead use a prune ratio $P$ as that is more representative and united across different environments: 
$$
P(C(G,\epsilon)) = \frac{\sum_{|g_i|<\epsilon}1}{n}.
$$
\subsubsection{Gradient noising}
Adding noise to gradients after training can help obscure individual contributions. For a gradient $G\in \mathbb{R^n}=(g_1, g_2, \dots, g_n)$, noised gradient is calculated as $$G_{noised} =(g_1, g_2, \dots, g_n)+(x_1, x_2, \dots, x_n),$$where $x_i$, with $1\leq i \leq n$, is generated from a distribution depending on type of noise, such as Gaussian noise used in this work.


\section{Related Work}

The method described by Jiang et al. \cite{GAIN} proposes a novel decentralized privacy-preserving FL scheme designed to address critical vulnerabilities in traditional FL, such as privacy leakage from gradient sharing and the single-point-of-failure problem. The scheme introduces pairwise masking combined with additively homomorphic encryption (AHE) to blind gradients, ensuring the confidentiality of participant data and resistance to quantity inference attacks. 

Khan et al. \cite{lovehatesplit} extend split learning by employing homomorphic encryption (HE) to protect activation maps exchanged during training. A fully-connected layer is chosen to be encrypted using HE for calculations on the server. HE allows computations to be performed directly on encrypted data, eliminating the need for decryption during processing. While HE offers robust privacy guarantees, its application 
is often constrained by high computational overhead.  

The study by Jia et al. \cite{chunkCNN} explores a privacy-preserving framework that combines homomorphic encryption with chunk-based convolutional neural networks. The proposed method uses a modified CNN that processes encrypted image chunks independently, improving efficiency. By selectively applying HE to image regions with high gradient levels, they achieve a better balance between security, speed and accuracy. The experimental results highlight a significant reduction in storage and transmission costs, along with improvements in classification accuracy, compared to conventional HE-based methods.


\begin{figure}
    \centering
    \includegraphics[width=0.95\columnwidth]{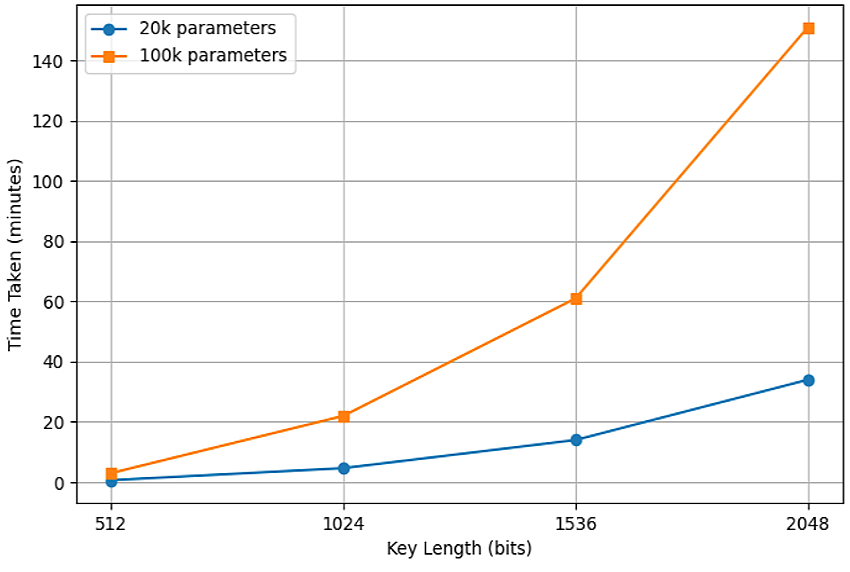}
    \caption{Time needed for different key lengths and parameter count in Paillier encryption.}
    \label{fig:pail-bad}
\end{figure}

\begin{figure}
    \centering 
    \includegraphics[width=0.96\columnwidth]{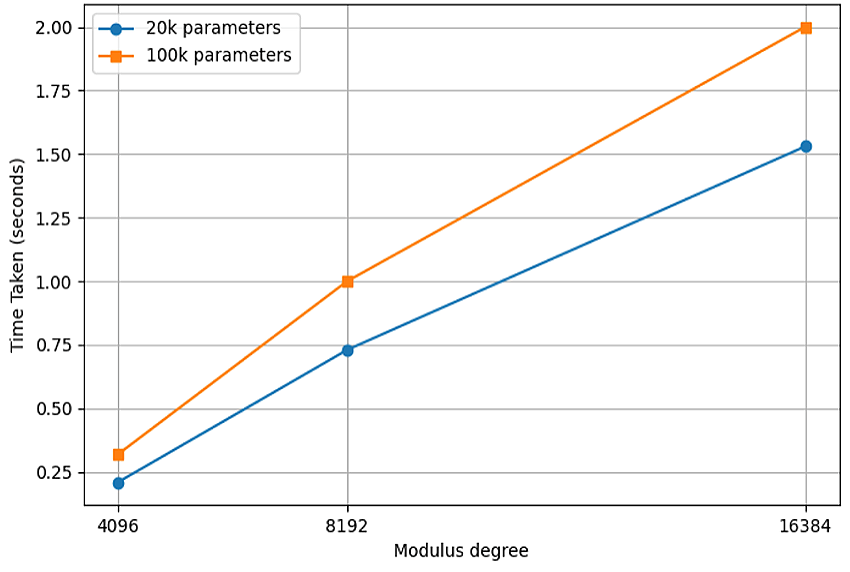}
    \caption{Time needed for different polynomial degrees and parameter count for CKKS encryption.}
    \label{fig:ckkstime}
\end{figure}

\section{Our Implementation}

\subsection{Server-side privacy}

In our experiments, we have used two Python libraries for the mentioned encryption schemes, namely the phe\footnote{https://pypi.org/project/phe/1.5.0/} library for Paillier and the TenSeal\footnote{https://pypi.org/project/tenseal/0.3.15/} library for CKKS.
A MacBook M3 Pro was used as hardware for these experiments. 

In Figure \ref{fig:pail-bad} we can see the impractical overhead of Paillier encryption. Since a 2048-bit key length corresponds to approximately 112 bits of security, which is the minimum standard for practical uses, along with the fact that much larger networks are used in practice, encrypting a model with 100,000 parameters for over 2 hours seems highly impractical. 


For CKKS, TenSeal offers a table of CKKS parameter values that are equivalent to 128 bits of security.
In our setup, for each polynomial degree, we adjust the modulus so that it's equivalent to approximately 128 bits of security. The scaling factor is set to $2^{40}$.


In Figure \ref{fig:ckkstime}, we can observe the drastic improvement in encryption times for different polynomial degrees.
Note that the $y$ axis is in seconds, while Figure \ref{fig:pail-bad} has the $y$ axis in minutes.
Although parameters are adjusted to maintain 128 bits of security, a higher polynomial degree enables us to increase the modulus, which in turn allows us to perform more computations before noise accumulation is significant. Further experiments have shown that noise caused by addition is insignificant\footnote{10000 additions caused an absolute error in the range of $10^{-14}$ with a scaling factor set to $2^{40}$.}, as it can be well managed using an appropriate scaling factor. 

\subsection{Client-side Privacy}

To evaluate the efficiency of data reconstruction and mitigation methods discussed previously, we conducted experiments using the DLG algorithm \cite{GLeakage}.

\begin{figure*}
    \centering
    \includegraphics[width=0.95\textwidth]{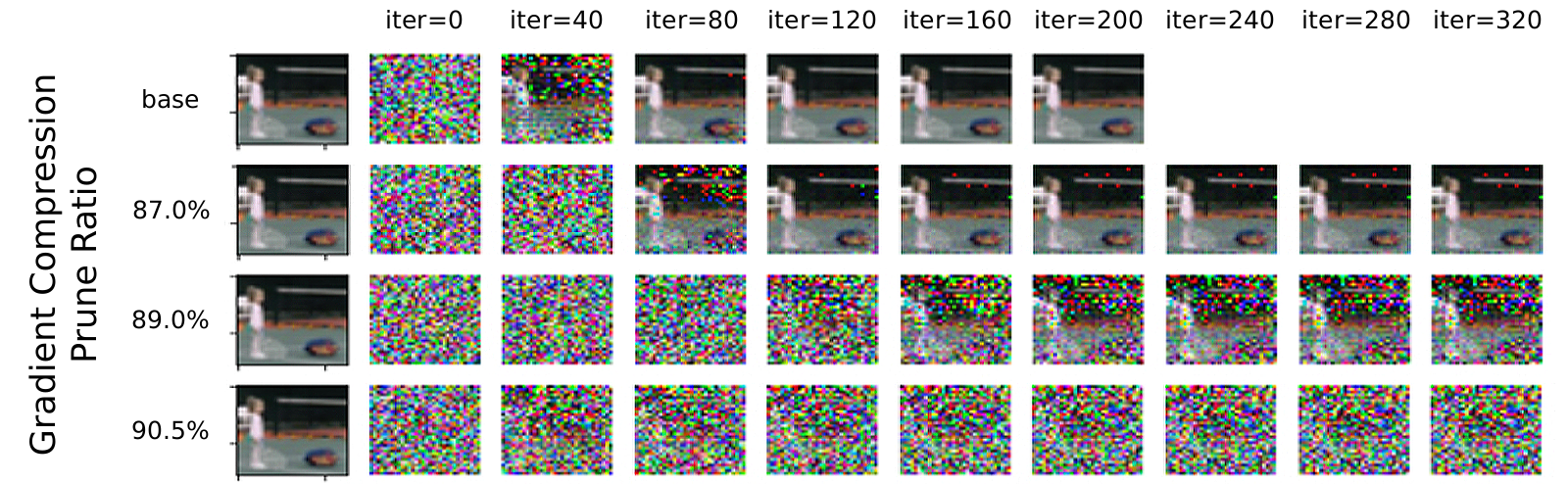}
    \caption{Reconstruction results for different prune ratios in gradient compression.}
    \label{fig:recGradPrune}
\end{figure*}

\begin{figure*}
    \centering
    \includegraphics[width=0.95\textwidth]{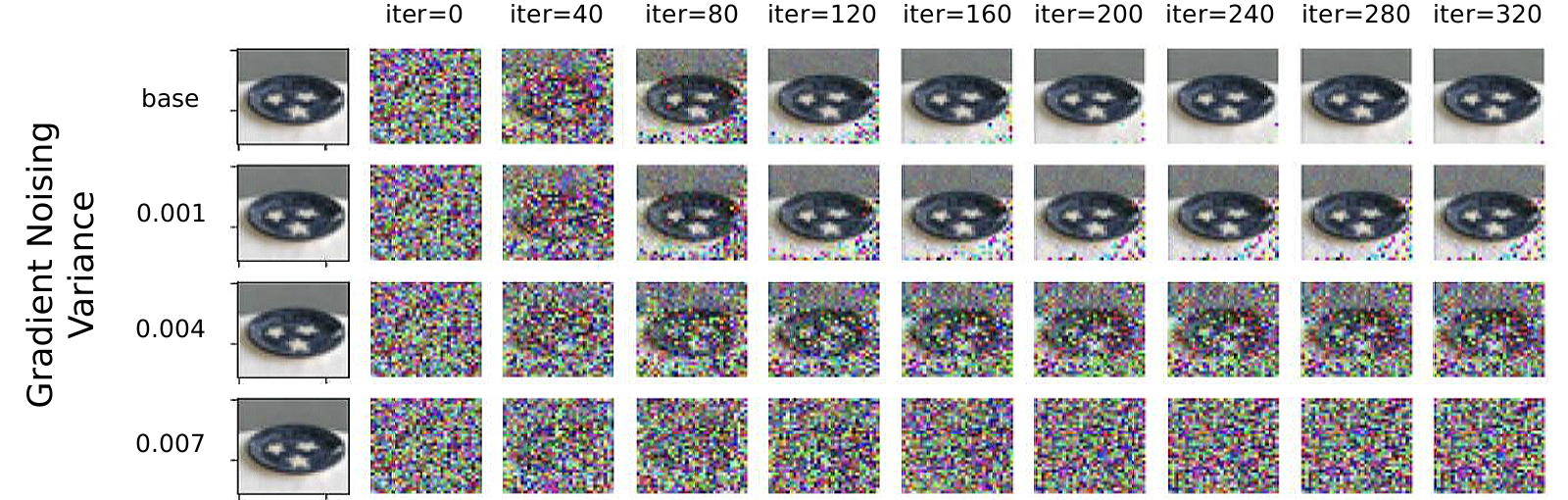}
    \caption{Reconstruction results for different variances in gradient noising (normal distribution).}
    \label{fig:gnoisingReconstruction}
\end{figure*}

\subsubsection{Classification network}
In this experiment, we employed a simple CNN designed for image classification. The network was implemented using the PyTorch\footnote{https://pypi.org/project/torch/2.7.0/}
library and consists of four layers. Specifically, the first three layers are convolutional layers, each with a kernel size of $5$ and utilizing a sigmoid activation function. The final layer is a fully connected layer, which employs a softmax activation function to classify input data into 10 distinct classes. 

The experimental data included input images and corresponding labels from the CIFAR-10 dataset \cite{cifar10}. All input images had dimensions of $32 \times 32$ pixels with three color channels 
The network parameters were initialized randomly from a normal distribution without any prior training. We used the cross-entropy loss function as a metric of error. 

For the optimization component of the DLG algorithm, we utilized the L-BFGS optimizer \cite{LBFGS} configured with a learning rate of 1 and max iterations set to $20$. Within the DLG procedure, the optimization objective is to minimize the sum of squared differences between the actual gradient and the reconstructed (dummy) gradient.

In Figure \ref{fig:recGradPrune}, we can see the result of reconstruction for different prune ratios using gradient compression. The base row had no compression. Up to an 87\% pruning ratio, compression had no noticeable effect, and the reconstructed images appear visually indistinguishable from the originals. The optimization process gradually worsens as prune ratio goes to $90.5\%$, where the result appears to become a random noise\footnote{While it appears that the results became a random noise, the actual pixel values, local neighborhoods and global statistics after optimization might carry critical information about the original image, for which there could exist mathematical techniques that would transform this seemingly random noise closer to the original. 
}.

\begin{figure}
    \centering
    \includegraphics[width=0.99\linewidth]{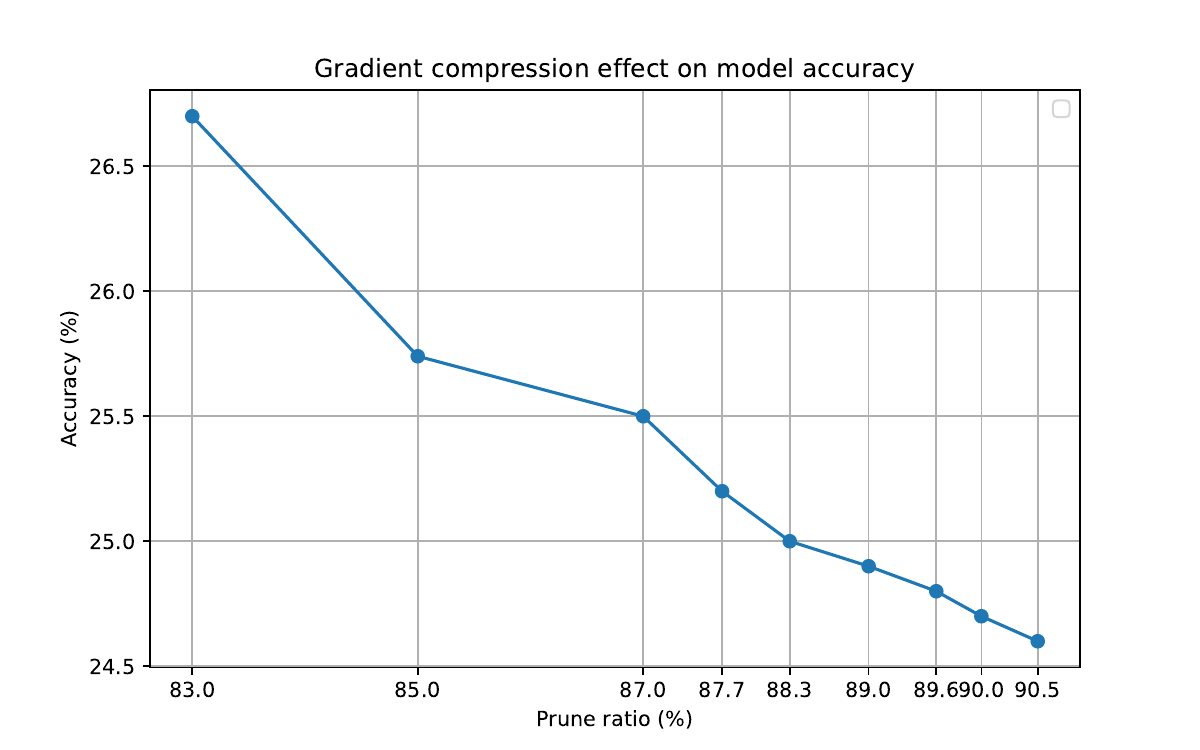}
    \caption{Negative effects of gradient compression on model accuracy.}
    \label{fig:GCompressionAccuracy}
\end{figure}

\begin{figure}
    \centering
    \includegraphics[width=0.99\linewidth]{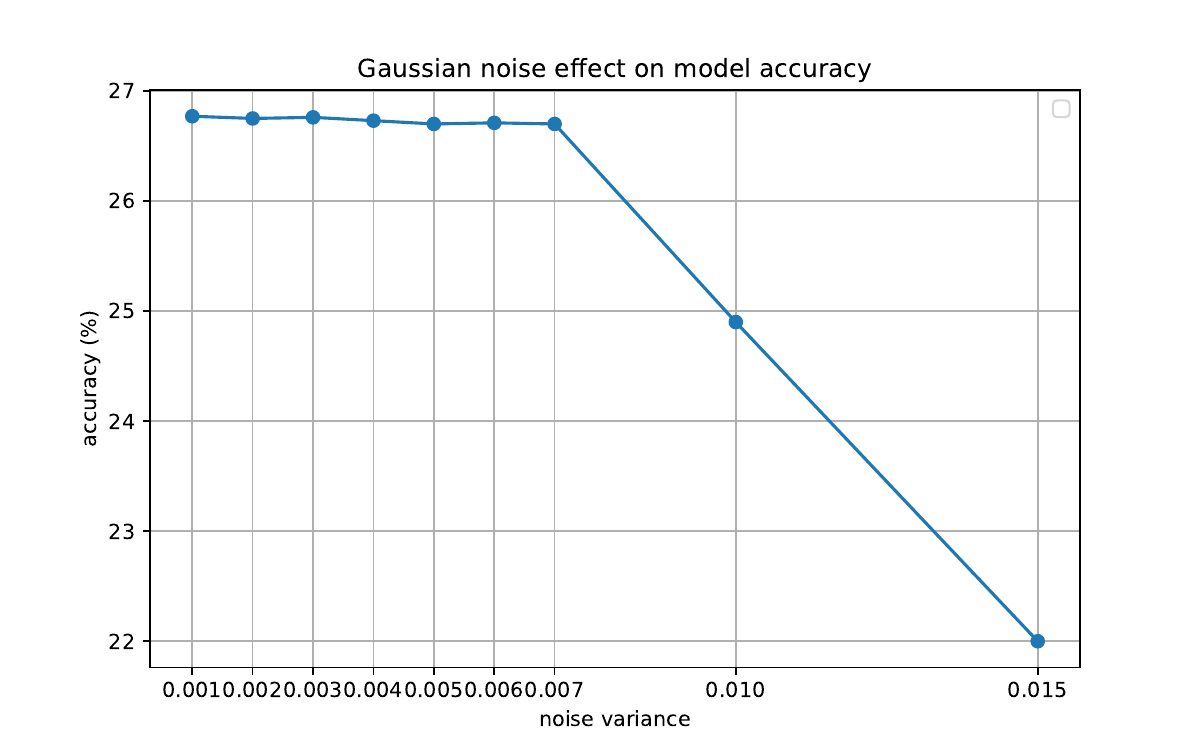}
    \caption{Negative effects of gradient noising (normal distribution) on model accuracy.}
    \label{fig:GNoisingAccuracy}
\end{figure}

Additionally, we implemented a local FL system using \texttt{OpenFL}\footnote{https://pypi.org/project/openfl/1.8/} framework \cite{openFL} and \texttt{NVIDIA FLARE}\footnote{https://pypi.org/project/nvflare/2.5.2/} library. 
In our setup, each client had the
same size of training sample data, while these sets were disjoint. Hardware-wise, the MacBook M3 Pro (used also in all the other experiments) was the server and also one of the clients, while the other clients were Jetson TX2\footnote{https://developer.nvidia.com/embedded/jetson-tx2} modules, representing the viability of this architecture on edge devices, such as mobile 3D scanners. While the MacBook took only 30 seconds to perform one training iteration (5 epochs), the Jetson TX2 required 5 minutes. 

We performed one FL iteration, where all clients were selected and merged, for different prune ratios to see the negative effect on the accuracy of the model after the aggregation round. This is displayed in Figure \ref{fig:GCompressionAccuracy}. Up to $83\%$ prune ratio, the model accuracy was not affected. As the prune ratio increased, the model's accuracy began to decline significantly earlier than the point at which reconstruction performance started to worsen. By the $87\%$ mark, 
the model already had a relative drop of $3.6\%$.  For the most significant compression, accuracy had a relative drop of $7.9\%$ in return for achieving visually good resistance against DLG image reconstruction.

We repeated the same process with gradient noising. In Figure \ref{fig:gnoisingReconstruction}, the top row starts with a variance of $0$ (no noising). Noising first starts impacting reconstruction at a variance of around $0.001$, which gets continuously worse until a variance of $0.007$, where the reconstruction, again, visibly becomes a random noise. What's more interesting is the impact of the noising on the same FL system as before, which can be seen in Figure \ref{fig:GNoisingAccuracy}. Gradient noising had no impact on the accuracy until the variance got near $0.007$, which is around the variance level where the reconstruction visibly fails. 
This makes gradient noising a viable candidate for protection against reconstruction relative to negative impact on model accuracy. 


\begin{figure*}[ht]
  \centering
  \begin{subfigure}{0.31\textwidth}
    \centering \includegraphics[width=0.95\columnwidth]{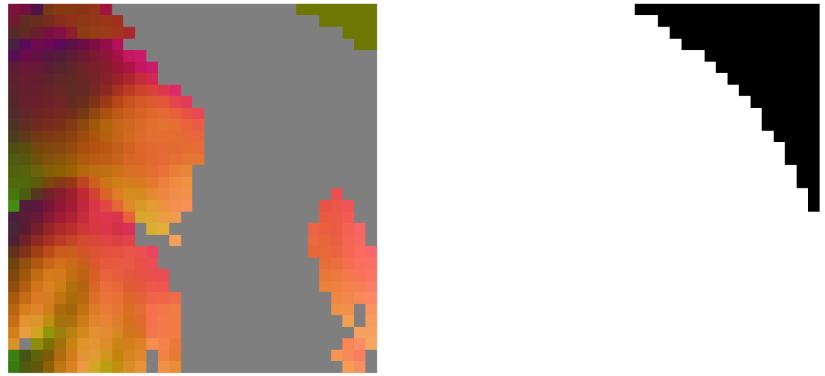}
    \caption{}
    \label{fig:gt-unet}
  \end{subfigure}
  \begin{subfigure}{0.31\textwidth}
    \centering \includegraphics[width=0.95\columnwidth]{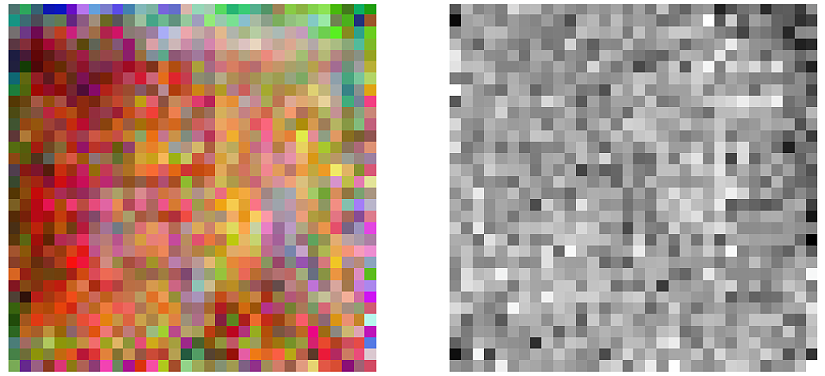}
    \caption{}
    \label{fig:seg-reconstruction}
  \end{subfigure}
  \begin{subfigure}{0.31\textwidth}
    \centering \includegraphics[width=0.95\columnwidth]{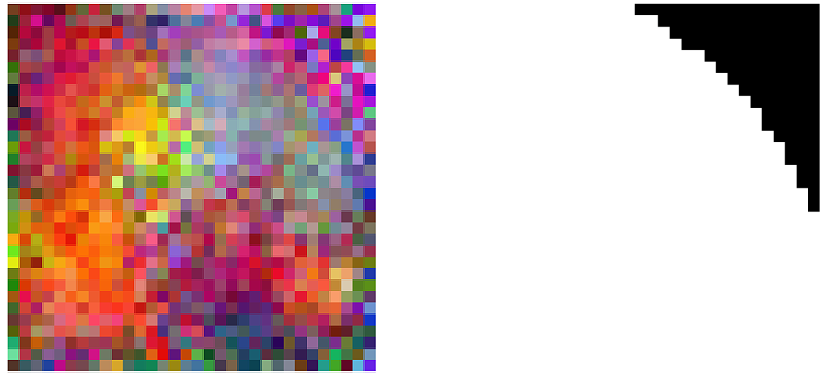}
    \caption{}
    \label{fig:seg-mask-reconstruction}
  \end{subfigure}

  \caption{
    (a) Ground-truth sample, (b) the reconstruction process result after 1200 iterations and (c) the result after 1200 iterations of input-only reconstruction. In each sub-figure, the normal map is displayed on the left and the segmentation mask on the right.}
  \label{firstnew}
\end{figure*}

\subsubsection{Segmentation network}
A similar experiment was conducted using a simplified U-Net segmentation network
. 
For input data, we have used cropped parts of a 3-channel normal map obtained from a 3D scanner.
The GT label consists of a single-channel binary mask of identical dimensions, representing a segmentation mask.
The network was initialized with random parameters obtained from a normal distribution. The loss function used for network training was binary cross-entropy. In the DLG algorithm, the objective was defined as the minimization of the sum of cosine similarities between the real and dummy gradients. 

The ground truth is displayed in Figure \ref{fig:gt-unet}. 
Figure \ref{fig:seg-reconstruction} displays the best result after 1200 iterations. While some aspects of the converged result are close, such as 
the color distributions in the normal map, overall result has failed to converge to a recognizable state. Neither gradient compression nor noising was used in this process. 
This process was repeated for several different inputs with different hyperparameter combinations\footnote{Various learning rates in interval $(1^{-6},1)$, penalty weights, different optimizers (AdamW, Adam, Adagrad, SGD, ...), DLG algorithm loss functions (mean squared error, cosine similarity).} and with up to 10 000 iterations. The displayed result is both visually and in terms of gradient cosine similarity, the closest result we managed to obtain. We think there might be two main reasons for this:
\begin{enumerate}
    \item In comparison with CIFAR-10 classification, where output was a vector of length 10, here the output is a whole image of size $30 \times 30$ pixels. The number of combinations is significantly larger, and increasing number of iterations did not help, as the optimization tends to diverge. 
    \item 
    Multiple input variants may compute similar gradient information.
    At the same time, there exist different versions of the input sharing the same mask. This can cause a large number of local minima and trouble for the optimisation to converge.
\end{enumerate}

To further elaborate on the first point, Figure \ref{fig:seg-mask-reconstruction} displays the result of the DLG algorithm when a real mask is given and only the input is optimised. While the results are not as close to the ground truth as in the case of a classification network,
the reconstructed normal map has much sharper edges, improved structure and color distribution compared to before.

\section{Conclusion}
In this work, we experimented with an FL framework tailored to mobile edge devices with limited resources, specifically using NVIDIA Jetson TX2 chips. We described theoretical mechanisms such as homomorphic encryption, client parameter aggregation, and gradient‐level privacy countermeasures. A software architecture for a local FL system was realized using standard Python libraries (TenSEAL, PyTorch/TensorFlow) and the OpenFL framework. 
Proof‑of‑concept deployment was demonstrated using CKKS encryption.

Through systematic experimentation, we quantified the trade‑offs between privacy and utility across multiple dimensions. On the server side, Paillier encryption, while secure, was shown to be impractically slow at recommended key sizes, with up to 2 hours for a 100000-parameter model, whereas the CKKS scheme achieved equivalent security, with orders-of-magnitude faster encryptions in a few seconds, and manageable noise growth under weighted averaging. On the client side, we compared two privacy-preserving techniques for protection against data reconstruction using the DLG algorithm, namely the gradient compression and gradient noising. While both have shown to be usable techniques for protection, experiments have shown that gradient noising increases protection for much lower costs in terms of lost model accuracy. Finally, segmentation network experiments revealed that reconstruction is significantly harder for complex networks and data resembling real-world scenarios, highlighting both the effectiveness and limits of current DLG methods.

Despite these successes, there is an important note. We have not provided formal guarantees that privacy is preserved, but instead looked at the (in)efficiency of data reconstruction using the DLG algorithm with proper privacy-preserving techniques in place. Different methods for data reconstruction may be more efficient. 

Open problems highlight the difficulty of determining when privacy is truly breached. Even if the reconstruction is not an exact copy, it may still reveal visual or structural information that is recognizable and sensitive. Numerical error measures used today often miss this, and future work should focus on practical metrics that reflect the actual information that an adversary could infer.

Architecturally, approaches such as split learning and fully encrypted models reduce the exposure of model parameters to either client or server, and therefore offer stronger privacy guarantees. However, they 
suffer from substantial computational overhead, especially on resource-constrained edge devices.

Going forward, a key direction is to combine more meaningful privacy metrics with lightweight secure architectures to move toward deployable and certifiable privacy-preserving federated learning in real-world 3D‐scanning scenarios.


\section{Acknowledgment}

This work was funded by the EU NextGenerationEU through the Recovery and Resilience Plan for Slovakia under the project ''InnovAIte Slovakia, Illuminating Pathways for AI-Driven Breakthroughs" No.~09I02-03-V01-00029.


\bibliographystyle{plain}
\bibliography{main} 

\end{document}